\documentclass[10pt,leqno]{amsart}
\usepackage{graphicx}
\usepackage{indentfirst,csquotes}

\usepackage{amssymb,amsthm,amsmath}
\usepackage{xcolor,paralist,hyperref,fancyhdr,etoolbox}

\hypersetup{ colorlinks=true, linkcolor=black, citecolor=black, filecolor=black, urlcolor=black }

\makeatletter
\renewcommand\paragraph{\@startsection{paragraph}{4}{\parindent}%
  {1.0ex plus .2ex minus .2ex}
  {-0.5em}
  {\normalfont\normalfont\bfseries}}
\makeatother

\usepackage{booktabs}
\usepackage{caption}   
\usepackage{tabularx}
\usepackage[labelfont=bf,labelsep=period]{caption} 

\usepackage{tikz}
\usetikzlibrary{positioning}
\usepackage{float}

\begin{document}
\title{Algorithmic Management and the Future of Human Work: Implications for Autonomy, Collaboration, and Innovation} 
\author[Konjen]{Huram Konjen}
\address{Nykredit, Copenhagen, Denmark}
\email{hurk@nykredit.dk}

\begin{abstract}
This study examines the evolving impact of algorithmic management on human resource management (HRM) practices, with a focus on employee autonomy, procedural transparency, and the sociotechnical dynamics of performance evaluation. Rather than adopting a qualitative or empirical approach, the paper develops a conceptual integration of insights from HRM, human-computer interaction (HCI), and Science and Technology Studies. The analysis highlights that although algorithmic systems can enhance operational efficiency, they risk reinforcing biases and narrowing the relational and contextual dimensions of work. These systems often overlook intangible contributions such as creativity, empathy, and collaborative problem solving, revealing gaps in data-driven performance measurement. In response, the study proposes a sociotechnical perspective on algorithmic accountability that emphasizes procedural transparency, organizational justice, and employee agency. By revisiting foundational questions within the rapidly evolving landscape of algorithmic management, the paper contributes to ongoing debates about the future of work and the design of managerial technologies that support, rather than constrain, human autonomy and organizational life.
\end{abstract} 

\maketitle


\bigskip

\section*{Preface}
This study originated in 2017, when algorithmic management was still an emerging phenomenon. At that time, adoption was uneven, and many organizations were reluctant to admit that they were using algorithmic systems for management, unless the narrative supported operational optimization or user empowerment. Access to empirical data was limited, and scholarly literature was just beginning to coalesce.

With eight years of hindsight and a wave of new empirical and theoretical research, this revised version returns to its original premises, updating them where necessary and reaffirming them where they hold. For example, Uber's case study remains included not because it is cutting-edge in 2025, but because it exemplifies one of the earliest and most studied instances of algorithmic management. As the field has matured, the concerns raised in this study have gained broader relevance across industries, providing a foundation for further critical and interdisciplinary inquiry into algorithmic management and its implications for the future of human work.

\section{Introduction}
The growing presence of intelligent algorithms in organizational contexts has fundamentally transformed management practices. In this study, intelligent systems are defined as algorithmically driven agents, ranging from rule-based logic to complex machine learning models, that perform decision-making functions previously undertaken by human managers. These systems are increasingly integrated into core business operations, including task allocation, scheduling, performance monitoring, and compensation.

Digital-native firms, particularly platform-based organizations such as Uber, Amazon and Google, have become emblematic of this shift \cite{Porter1995}. Unlike traditional organizations that later adopted digital tools, digital-native firms are those conceived around data-driven infrastructures from inception. Their core business models, databases, and processes are designed to leverage algorithmic decision-making and scale efficiently through machine learning models and automation technologies. These companies leverage algorithmic tools not only for operational optimization but also to manage human labor through automated systems \cite{Latour2005}. \textit{Algorithmic management} refers to the delegation of managerial tasks, such as work assignment, supervision, and evaluation, to software systems \cite{boyd2012}. This trend is underpinned by a techno-managerial belief that data abundance improves decision-making accuracy, fairness, and efficiency.

However, algorithmic management is more than just an efficiency upgrade; it reconfigures fundamental aspects of HRM. Algorithms determine how work is distributed, performance is assessed, and employees are incentivized or sanctioned \cite{Barocas2016}. For example, Uber’s dynamic pricing, task assignment, and performance scoring systems illustrate how managerial authority is embedded in algorithmic infrastructures that adjust labor allocation and compensation in real-time. While these systems can optimize operations, they simultaneously reshape work relations and challenge traditional HRM functions.

While Uber’s system exemplifies the efficiency gains of data-driven management systems, it also exposes its human costs, particularly the erosion of employee autonomy, fairness, and long-term work relations. These concerns extend beyond individual cases, reflecting a broader gap in existing research, which has largely emphasized technical performance and operational metrics while overlooking the sociocultural consequences for employees. This study addresses this gap by analyzing how algorithmic regimes reshape the employee experience, particularly in relation to autonomy, collaboration, and innovation within organizations. The analysis considers how these systems influence workplace practices beyond technical outcomes by engaging with the sociotechnical dynamics they introduce.

To clarify how these dimensions are understood in this study, Table \ref{tab:keyconstructs} summarizes the core constructs, such as autonomy, collaboration, and innovation, and how they are typically conceptualized in sociotechnical research.

\subsection*{Key Constructs and Conceptual Definitions}
\label{tab:keyconstructs}

\setlength{\tabcolsep}{6pt}
\renewcommand{\arraystretch}{1.25}

\noindent
\vspace{0.3\baselineskip}

\noindent
\begin{tabular*}{\linewidth}{@{\extracolsep{\fill}}p{.14\linewidth} p{.80\linewidth}}
\toprule
\textbf{Construct} & \textbf{Conceptual Definition and Illustrative Indicators} \\
\midrule
\textbf{Autonomy} & Degree of employee discretion in deciding \textit{how}, \textit{when}, and \textit{what} tasks are performed. 
Observable through decision latitude, scheduling freedom, and capacity to deviate from algorithmic recommendations. \\[3pt]
\textbf{Collaboration} & Quality and density of interactions between employees and teams in achieving shared goals. 
Can be reflected in reciprocity of communication, peer coordination, and participation in collective problem solving. \\[3pt]
\textbf{Innovation} & Generation and implementation of new ideas or process improvements within the organization. 
Observable through employee proposals, experimentation, and adoption rates of novel solutions. \\
\bottomrule
\end{tabular*}

\vspace{-6pt}
\captionof{table}{Core constructs and illustrative indicators.}
\vspace{4pt}
\noindent{\scriptsize \textit{Definitions are conceptual anchors drawn from interdisciplinary HRM, HCI, and organizational behavior literature, reflecting how these constructs are typically operationalized in sociotechnical research.}}
\vspace{10pt}

\newpage
These constructs form the conceptual basis for the subsequent analysis of how algorithmic regimes affect employees and organizations. This evolution raises critical questions about the human experience under algorithmic regimes. Employees increasingly interact with opaque, impersonal decision-makers who lack the capacity for discretion, empathy, and contextual judgment \cite{Burrell2016, Pasquale2015}. As Pasquale argues in The Black Box Society, algorithmic opacity concentrates informational power in organizations while concealing decision logic from those it governs, eroding transparency and accountability. Concerns include diminished autonomy, unclear evaluation criteria, and psychological effects of working under constant algorithmic surveillance. The replacement of human discretion with automated oversight challenges long-standing notions of organizational justice, fairness, and accountability.

Accordingly, this study poses the central research question: \textit{How do employees perceive and respond to the shift from human managers to algorithmic management?} In doing so, the study analyzes the impact of intelligent systems on employee autonomy, collaboration, and organizational innovation. Rather than assessing the technical functioning of algorithms, this work foregrounds the implications of these systems for workplace practices and organizational life.

This study considers how intelligent managerial systems reshape task execution, creativity, and employee autonomy, influencing how work is allocated and innovation is fostered across organizational structures. The growing reliance on such systems not only transforms operational workflows but also the sociotechnical conditions that enable collaboration and innovation. By drawing on research on the lived experience of algorithmic supervision, this work contributes to broader debates on the future of work under digital governance.

This article is a conceptual integration paper that synthesizes insights across HRM, HCI, and STS to develop an analytical framework for understanding how algorithmic management reshapes autonomy, collaboration, and innovation.

This study contributes by (1) synthesizing sociotechnical perspectives into an integrative framework for understanding algorithmic management, and (2) proposing three avenues for future research and design.

Distinct from prior work that has examined these domains separately, this study unites them into a coherent model linking employee autonomy, collaboration, and innovation under algorithmic governance. By translating conceptual tensions into actionable design and governance principles, it offers a compact pattern catalog for responsible algorithmic management, bridging theoretical insight and practical application. 

\begin{figure}[H]
\centering
\begin{tikzpicture}[scale=5, font=\scriptsize]

\coordinate (A) at (0,0);
\coordinate (B) at (1,0);
\coordinate (C) at (1,1);
\coordinate (D) at (0,1);

\fill[gray!5!white] (A) -- (C) -- (D) -- cycle; 
\fill[green!10!white] (A) -- (B) -- (C) -- cycle; 
\draw[line width=0.5pt, color=gray!60, rounded corners=0.05cm] (A) -- (B) -- (C) -- (D) -- cycle;
\draw[line width=0.5pt, color=gray!60] (A) -- (C);

\node[align=center, anchor=east, xshift=-0.04cm] at (0,0.75) {Standardization\\vs.\ discretion};
\node[align=center, anchor=west, xshift=0.04cm] at (1,0.75) {Surveillance\\vs.\ trust};
\node[align=center, anchor=east, xshift=-0.04cm] at (0,0.18) {Metric tightness\\vs.\ exploration};
\node[align=center, anchor=west, xshift=0.04cm] at (1,0.18) {Efficiency\\vs.\ well-being};

\node[font=\footnotesize\bfseries, above=6pt of C] {System goals};
\node[font=\footnotesize\bfseries, below=6pt of A] {Human goals};

\node[rotate=45, align=center, font=\scriptsize\itshape] at (0.5,0.5)
{Sociotechnical balance zone\\
(transparency, participation, appeal, oversight)};

\node[font=\small\bfseries, above=16pt of D] {Conjoined Triangles of Algorithmic Tension.};

\end{tikzpicture}

\caption{A visual metaphor illustrating tensions between system-level efficiency and norms associated with organizational justice and employee autonomy. The diagonal zone represents the balance area where governance mechanisms mediate these competing goals. Each tension captures a recurring trade-off documented in HCI and HRM studies of algorithmic management. The model is intended as a descriptive map of competing imperatives rather than a prescriptive solution.}
\label{fig:conjoined-triangles}
\end{figure}
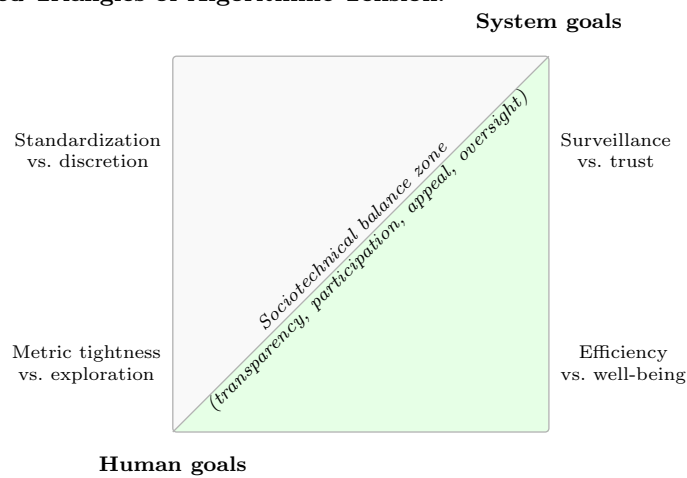

These tensions are synthesized visually in Figure~\ref{fig:conjoined-triangles}, which conceptually maps the dual demands of algorithmic management on both system and human goals. The model highlights the balance between efficiency-oriented system imperatives and norms of organizational justice, including discretion, trust, and employee well-being.

\newpage
\section{Theoretical and Conceptual Foundations}
An interdisciplinary theoretical approach grounded in HRM, HCI, and Science and Technology Studies is adopted to understand how algorithmic systems reshape employee experience. Traditional HRM practices, centered on human judgment, qualitative evaluation, and interpersonal feedback, have evolved significantly with the advent of algorithmic management. Historically, managers relied on experiential and social knowledge to evaluate performance and allocate tasks \cite{Guest1997}, whereas algorithmic systems now perform these functions through quantifiable metrics and automated assessments.

In contrast to traditional managerial judgment, algorithmic systems assess performance based on productivity, task completion times, and behavioral data \cite{BrynjolfssonMcAfee2014}. These systems promise greater consistency and scalability but raise critical questions about transparency, fairness, and the erosion of human discretion. Understanding these tensions requires integrating insights from HRM, HCI, as well as Science and Technology Studies to analyze how algorithms function not merely as tools, but as active agents shaping organizational life.

From an HRM perspective, this study analyzes how digital systems mediate managerial functions such as evaluation, feedback, and incentive structures. HCI provides insights into how algorithmic interfaces shape worker interactions with digital tools, emphasizing the affordances and constraints that influence behavior. The paradoxes in algorithmic HRM, particularly the challenges posed by technology affordances in HRM, are explored in recent studies \cite{ZouZhou2024}. Science and Technology Studies, especially Actor-Network Theory (ANT) \cite{Latour2005} and Foucauldian theories of surveillance, offer tools to conceptualize how algorithms redistribute power, structure visibility, and reconfigure agency in the workplace.

Rather than treating these perspectives as isolated, this study integrates them to analyze algorithmic regimes as a complex sociotechnical system. This approach moves beyond evaluating algorithmic efficiency to consider its cultural, ethical, and relational effects on work environments. As Kellogg, Valentine, and Christin \cite{Kellogg2020} highlight, algorithmic control constitutes a new terrain of contestation in organizations, shifting how authority, visibility, and decision rights are distributed between humans and machines. By combining these frameworks, the analysis provides a deeper understanding of the dual role of algorithms as both technical tools and active agents within organizational systems.

While much existing literature focuses on the operational efficiencies of digital managerial control, few studies have explored the sociocultural impacts of these systems on employee experience, autonomy, and organizational dynamics, which this study addresses. Building on these theoretical foundations, this analysis highlights how data-driven management practices influence autonomy and organizational culture, areas that remain underexplored in current research.

\subsection{Socio-Technical Systems Theory (STS)}
STS emphasizes the interdependence between technology and the social arrangements within which they are embedded. Rather than viewing technologies as neutral tools, this perspective considers how algorithmic systems co-constitute organizational norms, workflows, and relationships. In the context of algorithmic management, this theory highlights how task coordination, communication practices, and innovation processes are shaped by technological integration \cite{Zhang2025}.

By acknowledging the interdependent nature of technology and society, STS underscores the importance of considering both technical and social dimensions when designing and implementing such systems. This approach serves as a reminder that the influence of algorithms extends beyond their technical functions, shaping organizational cultures and employee experiences.

\subsection{Surveillance and the Panopticon}
To understand how power operates in algorithmic systems, this study draws on Michel Foucault’s concept of \textit{panopticism}. Algorithms enable continuous, data-intensive, and opaque forms of digital surveillance. Workers are increasingly subjected to real-time tracking, behavior scoring, and predictive assessments that operate invisibly and without recourse. This dynamic parallels what Zuboff describes as surveillance capitalism, a regime in which data extraction becomes a mechanism of behavioral control \cite{Zuboff2019}. Such oversight fosters a climate of internalized control, where the anticipation of observation becomes a mode of governance \cite{Foucault1977}. Algorithmic panopticism is particularly insidious because it is often perceived as neutral and objective, masking asymmetries in visibility, influence, and accountability.

Foucault’s concept of surveillance highlights how algorithmic systems redefine power dynamics in the workplace. These systems shift authority from human supervisors to algorithms, creating new forms of control that affect employee behavior and perceptions of fairness. This shift has profound implications for organizational justice and employee trust.

Recent studies on algorithmic surveillance in the gig economy further illustrate how digital platforms reorganize work through spatial and behavioral control \cite{Newlands2021}. Newlands shows how the physical and virtual spaces in which workers operate are increasingly shaped by automated evaluation mechanisms.

While prisons have largely abandoned Panopticon-based designs due to ethical concerns, modern workplaces have begun adopting similar principles under the guise of efficiency and optimization. The irony lies in the fact that, even as penal systems move toward rehabilitation, organizations embrace algorithmic oversight that normalizes constant observation and control.

This development raises well-documented ethical concerns regarding autonomy, privacy, and the human cost of digital surveillance. Just as the Panopticon redefined power dynamics within prisons, algorithmic oversight is profoundly reshaping the relationship between employees and employers, often through invisible and unaccountable mechanisms of control. Ensuring that these systems respect transparency, fairness, and dignity is essential if algorithmic management is to serve rather than subjugate the modern workforce.

\subsection{Actor-Network Theory}
ANT offers a view of agency that recognizes technological artifacts, such as recommendation engines or dashboards, as active participants in networks of organizational control \cite{Latour2005}. In algorithmic management, ANT helps reveal how algorithms are not merely tools but actors that shape decisions, mediate communication, and influence outcomes. This perspective enables a tracing of how managerial authority is distributed across humans and machines in hybrid assemblages that challenge traditional notions of hierarchy and responsibility.

By viewing algorithms as actors, ANT encourages a more dynamic understanding of organizational power. This prompts a reconsideration of how authority is shared between human and non-human agents, emphasizing the importance of relational dynamics in algorithmic systems. This perspective is particularly useful in exploring how algorithmic oversight reshapes leadership structures and decision-making processes.

\subsection{Algorithmic Accountability}
As algorithms assume managerial roles, the need for robust accountability mechanisms becomes increasingly urgent. The concept of algorithmic accountability, which emphasizes transparency, explainability, and oversight, is essential for ensuring fairness in algorithmic decision-making \cite{Crawford2021, Barocas2016}. In HRM contexts, these concerns highlight the risk of bias, loss of trust, and limited recourse for employees, making algorithmic accountability crucial for ensuring fairness. Algorithmic accountability aims to introduce governance structures, both technical and organizational, that ensure that algorithms operate in accordance with ethical and legal standards \cite{Latour2005}.

Algorithmic accountability connects directly to the ethical and legal challenges that arise as algorithmic governance becomes embedded in routine organizational workflows. This underscores the need for clear frameworks that protect employees from unfair practices and ensure transparency in algorithmic decision-making. This focus on accountability provides a critical lens for examining the long-term consequences of algorithmic management on organizational integrity and employee trust.

\subsection{Bringing the Approaches Together}
Combined, these approaches enable a layered understanding of algorithmic management as a sociotechnical phenomenon. STS situates algorithmic systems within organizational structures, revealing how they shape and are shaped by social dynamics. Foucauldian surveillance theory and ANT explore how power and agency are redefined in algorithmic regimes, demonstrating how algorithms mediate relationships and influence decision-making. Algorithmic accountability provides a critical lens for assessing the ethical and legal challenges presented by these systems, ensuring that they align with principles of fairness and transparency.

Together, these frameworks illustrate that algorithmic systems cannot be understood in purely technical or managerial terms. It operates through intertwined social and technological processes that redefine how authority, autonomy, and collaboration are enacted in organizations. Recognizing this interdependence sets the stage for examining the broader implications of algorithmic oversight in shaping the future of work.

\section{Discussion}
Algorithmic management represents a fundamental reconfiguration of managerial authority, shifting key HRM functions, such as task allocation, performance evaluation, and reward systems, away from human discretion and toward data-driven systems. This shift brings both efficiencies and profound organizational challenges, and recent studies suggest that these systems are moving from an edge case to a more mainstream practice \cite{KeeganMeijerink2025}. Yet recent scholarship has questioned some of the core assumptions underlying this field, particularly the tendency to overlook worker experience and the organizational dynamics in the pursuit of automation \cite{Lamers2024}. Drawing from an interdisciplinary approach, this discussion analyzes these dynamics across four critical dimensions: the reduction of individuals to data points, the impact on employee autonomy and agency, the influence on collaboration, and the implications for innovation in the workplace.

The development of the Algorithmic Management Questionnaire \cite{ParentRocheleau2024} represents an important step toward quantifying and understanding the employee experience under algorithmic regimes and gaining insights into employees’ perspectives on algorithmic oversight.

The integration of STS and strategic human resource development \cite{Zhang2025} addresses some of the challenges identified in algorithmic systems, helping to better understand the complex interplay between technology and organizational dynamics.

\subsection{Reducing Individuals to Numbers}
A central concern of algorithmic management is its reductive treatment of workers, transforming complex human attributes into standardized metrics. This quantification facilitates scalability and consistency but does so by abstracting away from the nuanced qualitative aspects of employee performance and social interaction. The overemphasis on efficiency in algorithmic systems \cite{Lamers2024} often leads to neglecting relational dynamics and subjective worker experiences.

The problem is twofold. First, some forms of labor, such as mentoring, informal leadership, or creative contributions, are difficult to measure and are thus excluded from algorithmic evaluations. Second, even when data is available, algorithms often lack the contextual sensitivity to interpret it meaningfully. This introduces systematic blind spots, where employees who contribute value through unquantifiable means risk being undervalued or misrepresented in performance evaluations.

To mitigate these effects, organizations must explore hybrid approaches that integrate qualitative signals into algorithmic systems. Advances in large language models make it possible to incorporate peer feedback, team climate, and communication quality into decision-making processes in ways that earlier natural language processing techniques could not. Recent approaches to fair algorithms increasingly focus on explainability, human-in-the-loop design, and bias auditing mechanisms that ensure algorithmic decisions remain accountable to human oversight \cite{Lynn2025}. However, these solutions must be accompanied by institutional mechanisms for contestability, where employees can challenge and contextualize the data. Transparency, auditability, and participation are essential for maintaining employee trust and ensuring fair algorithms.

In summary, while data-driven decision-making offers operational advantages, overreliance on numerical abstraction risks eroding the relational and affective dimensions that sustain organizational life. Responsible algorithmic oversight requires a commitment to capturing the full spectrum of human contributions, especially those dimensions that are not easily rendered in data.

\subsection{Impact on Employees}
Algorithmic management shifts control toward algorithms, reducing employee autonomy and centralizing decision-making, often at the expense of human input. While some workers appreciate the perceived objectivity of algorithmic systems, many report a diminished sense of control and growing uncertainty regarding performance expectations. The removal of human interaction from supervisory processes reduces opportunities for feedback, mentoring, and empathetic engagement, which are key elements for employee development and well-being.

Policy evidence from an OECD employer survey underscores the global diffusion of algorithmic management and its accompanying regulatory challenges \cite{OECD2025}. These findings align with recent studies noting persistent difficulties in reconciling software-driven management systems with existing legal frameworks \cite{Lynn2025}.

The case of Uber drivers illustrates these dynamics \cite{Lee2015}. Uber was among the first firms to experiment with algorithmic labor control. Rosenblat and Stark’s ethnographic analysis reveals how information asymmetries, where the platform controls data on pricing, ratings, and performance, reinforce managerial authority and limit driver autonomy \cite{RosenblatStark2016}. In response to early backlash, Uber adjusted its algorithms to factor in rider-driver distance and decision time. This evolution demonstrates enduring trade-offs between efficiency and human cost as similar systems spread across industries \cite{KeeganMeijerink2025}. Comparable trends appear in other gig economy platforms, where intelligent systems regulate workers’ tasks, evaluation, and compensation, highlighting the widespread influence of algorithmic control \cite{Duggan2023}.

Beyond the gig economy, algorithmic oversight now shapes conventional workplaces. In warehouses, performance-tracking systems monitor pick rates and flag deviations automatically. Call centers use sentiment analysis and routing tools to allocate cases and assess service tone, while banks deploy triage algorithms to prioritize compliance reviews. Across these contexts, data-driven supervision has moved from the platform economy into mainstream management, reshaping autonomy, accountability, and everyday decision-making \cite{KeeganMeijerink2025,Stark2024}. This diffusion underscores a structural shift: algorithmic oversight has evolved from a platform innovation into a generalized management logic.

A consulting project at an insurance company similarly revealed the ethical challenges of algorithmic oversight. An algorithm introduced to reduce case backlogs improved efficiency but significantly reduced satisfaction among senior attorneys, who valued discretion in case selection. The episode illustrates the risks of siloed decision-making, where operational goals outweigh employee well-being, and underscores the need for governance mechanisms that balance algorithmic efficiency with human judgment.

Such systems introduce uncertainty and anxiety among workers. In response, employees exercise digital agency by organizing in online forums, sharing system insights, and collectively negotiating work conditions to reclaim control through informal, peer-driven networks. 
Outside the platform context, retail environments exhibit comparable dynamics. Public discourse reflects growing concern about pervasive tracking and timing systems that intensify managerial control over service work \cite{Floreani2024}. Meanwhile, the spread of algorithmic surveillance, particularly in gig work, constrains autonomy and transforms previously flexible spaces into sites of intensified control \cite{Newlands2021}. These dynamics suggest that algorithmic systems do not eliminate human agency but reshape it. For HRM, this underscores the importance of designing systems that are not only efficient but also adaptable, dialogical, and inclusive of employee feedback and agency throughout decision-making processes.

This growing complexity highlights the need for more nuanced HRM frameworks that balance technological efficiency with employee well-being and organizational trust. These practical implications are summarized in Table \ref{tab:implications}, which links observed risks to design and governance responses, demonstrating how sociotechnical tensions can be addressed through interdisciplinary management practices. As Mittelstadt et al. \cite{Mittelstadt2016} emphasize, effective algorithmic governance depends on aligning system design with principles of accountability, transparency, and human oversight, conditions essential for sustaining trust and fairness in digital workplaces.

\subsection{Impact on Collaboration}
The effects of algorithmic management are not limited to individuals; they ripple across teams and collaborative practices. By optimizing efficiency, algorithms influence collaboration by redistributing tasks and scheduling interactions, thus altering team dynamics and potentially reducing the autonomy traditionally associated with human decision-making. While this may enhance operational throughput, it can disrupt the social cohesion, trust, and informal communication that underpin effective collaboration in the workplace.

Traditional team formation processes foster interpersonal chemistry and psychological safety, whereas algorithmic optimization typically prioritizes task fit over relational depth, potentially weakening collaborative cohesion. Moreover, workers may hesitate to share information or take interpersonal risks if they believe such actions are not captured or rewarded by the system.

To counteract these trends, algorithmic system designers must consider the social dimensions of teamwork. Algorithms should not only be optimized for task allocation but also support relational processes by fostering transparency, equity, and shared goals. In practice, such systems are typically developed by software-as-a-service providers or internal data science teams within large organizations, whose design priorities often emphasize measurable productivity gains over relational or cultural factors. Incorporating metrics of team trust or communication quality, alongside productivity indicators, may help create systems that enable rather than constrain collaboration.

\subsection{Impact on Innovation}
While innovation often depends on slack, ambiguity and informal experimentation, algorithmic management can limit the flexibility needed for breakthrough ideas by standardizing tasks and optimizing processes. These conditions stand in tension with the efficiency-driven logic of these systems, which prioritize standardization and predictive control. Although data-driven systems can support incremental innovation by surfacing trends or optimizing processes, they often struggle to support radical innovation, which depends on informal experimentation and autonomous discovery.

By codifying what counts as “productive” behavior, algorithms may inadvertently discourage risk-taking or lateral thinking. Workers may optimize the metrics on which they are evaluated, leading to conformity rather than creativity. In addition, the reduced unstructured interaction time, often displaced by algorithmically scheduled meetings and rigid workflows, removes the contexts where novel or exploratory ideas typically emerge. In contrast, several innovation-driven firms deliberately balance algorithmic efficiency with human discretion. Organizations such as Pixar and IDEO rely on participatory, human-centered decision processes, while hybrid models at companies such as Spotify or Netflix combine data analytics with human curation to preserve creativity and contextual judgment.

Sustaining innovation under algorithmic regimes requires intentional design interventions. Organizations must carve out spaces for play, experimentation, and deviance from the algorithmic script. This may include flexible time structures, cross-functional hackathons, or human oversight roles that protect creative slack. Ultimately, innovation requires not only intelligent systems but also intelligent cultures, those that balance data-driven guidance with the messy, improvisational nature of human creativity.

\section{Implications for Research and Practice}

\paragraph{Design Considerations:}
The analysis emphasizes the need for intelligent systems that are not only technically efficient but also ethically aligned with human values. HCI designers should embed procedural transparency, contestability, and meaningful explainability into algorithmic management tools to ensure that users can understand and contest automated decisions. For example, integrating explainability features in such tools can provide employees with insights into decision-making processes. Ethical-by-design approaches are essential for building trust in systems that affect employment outcomes.

\paragraph{Participatory Design:}
This study reinforces the value of participatory design in the development of algorithmic management systems. Involving HR professionals, employees and other stakeholders as co-designers helps ensure that these technologies reflect the complexities of workplace contexts, including power dynamics, emotional labor, and informal collaboration. For example, co-design workshops could enable all parties to actively contribute to the system's design. HCI practitioners are urged to move beyond usability concerns toward systems that meaningfully support sociotechnical work practices.

\paragraph{Strategic Integration:}
HR professionals should guide and constrain the integration of algorithmic systems into core practices. Rather than simply adopting tools designed elsewhere, HRM should guide the implementation of intelligent systems to complement human judgment, contextual understanding, and ethical decision-making. This requires new competencies in data literacy and algorithmic governance.

\paragraph{Policy and Governance:}
Algorithmic management requires robust HR policies to govern the use of employee data, mitigate bias, and ensure fairness. These policies should establish clear standards for algorithmic accountability, including provisions for audit trails, human oversight, and employee recourse mechanisms. HRM must evolve into a guardian of technology integration and workplace justice.

\paragraph{Cultural Transformation:}
Algorithmic management introduces new tools and logics for control and evaluation, often reshaping not only organizational culture but also the physical and virtual spaces in which employees operate. To adapt, organizations are expected to cultivate a culture of openness, adaptability, and continuous learning that supports procedural transparency and responsible technology use.

\paragraph{The Future of Work:}
As intelligent systems become increasingly embedded in organizational life, stakeholders across industry, education, and policy must collaborate to prepare the workforce. Reskilling initiatives should address both the technical skills required to interact with these systems, and the socioemotional competencies needed to thrive in algorithmically mediated environments. Partnerships between businesses and educational institutions could play a key role in ensuring that workers are prepared for the evolving landscape.

\paragraph{Ethical Responsibility:}
The ethical implications of algorithmic management require ongoing scrutiny and dialogue. Organizations should engage in interdisciplinary research, regulatory developments, and employee perspectives to develop frameworks that ensure these technologies support autonomy, agency, and social equity. Ethical considerations must be integral to the design, development, and deployment of algorithmic systems. This includes establishing ethical guidelines for algorithmic decision-making processes and ensuring that organizations remain accountable to both employees and regulatory bodies.

\subsection*{Practical Implications Summary}
\vspace{0.5\baselineskip}

\setlength{\tabcolsep}{6pt}
\renewcommand{\arraystretch}{1.25}

\noindent
\vspace{0.3\baselineskip}

{\scriptsize
\noindent
\begin{tabular*}{\linewidth}{@{\extracolsep{\fill}} p{.20\linewidth} p{.21\linewidth} p{.30\linewidth} p{.20\linewidth}}
\toprule
\textbf{Risk or Tension} & \textbf{Observable Symptom} & \textbf{Design / Governance Response} & \textbf{Indicative Metric} \\
\midrule
\addlinespace[2pt]
\multicolumn{4}{l}{\textit{System Design Risks}} \\
Opacity
  & “Black-box” complaints
  & Explanation layers
  & Clarity score \\
Employee exclusion
  & Low adoption; resistance
  & Participatory design
  & Participation rate \\
Data-driven bias
  & Bias or appeal spikes
  & Bias audits + appeal workflow
  & Reversal rate \\
\addlinespace[4pt]
\multicolumn{4}{l}{\textit{Behavioral and Creative Constraints}} \\
Overreliance on metrics
  & Reduced initiative
  & Pair KPIs with peer feedback
  & Collaboration index \\
Autonomy loss
  & Creativity drops
  & Increase decision latitude
  & Autonomy index \\
\addlinespace[4pt]
\multicolumn{4}{l}{\textit{Cultural and Organizational Outcomes}} \\
Erosion of trust
  & Declining trust survey
  & Algorithmic literacy
  & Trust index \\
Transparency
  & Communication gaps
  & Leadership communication
  & Communication index \\
Efficiency focus
  & Burnout; high turnover
  & Balance KPIs with well-being
  & Retention; well-being \\
\bottomrule
\end{tabular*}
}

\vspace{-6pt}
{%
  \captionsetup{type=table,font=footnotesize} 
  \captionof{table}{Checklist linking risks to symptoms, responses, and metrics.}
  \label{tab:implications}
  \vspace{4pt}
  \begin{minipage}{\linewidth}\footnotesize
    \textit{This summary translates the study’s conceptual insights into actionable guidance for researchers and practitioners. It highlights how sociotechnical risks can be mitigated through design and governance responses that align system-level performance objectives with transparency, procedural fairness, and employee autonomy. Rather than prescribing fixed solutions, the table illustrates how interdisciplinary principles from HCI, HRM, and organizational theory can inform responsible algorithmic management practices.}  
  \end{minipage}
}

\section{Conclusion}
This study analyzed the intersection of HCI and HRM through the lens of algorithmic management. Drawing on an interdisciplinary approach, the study considers how intelligent systems are reshaping core organizational processes, particularly in relation to employee evaluation, autonomy, collaboration, and innovation.

This analysis highlights the dual nature of algorithmic oversight. On the one hand, it promises greater efficiency, consistency, and data-driven insights. However, this raises significant concerns about transparency, fairness, and the reduction of human complexity to quantifiable metrics. These tensions reflect deeper sociotechnical dynamics and underscore the need for systems that recognize both organizational benefits and human consequences of automation. Some of the assumptions underlying research in this domain have been critiqued \cite{Lamers2024}, urging a more balanced approach that considers both the technological potential and the human cost of these systems.

The discussion of quantification illustrates the limitations of algorithmic abstraction, particularly in capturing the relational, emotional, and contextual dimensions of human work. By relying on granular behavioral metrics such as response times, movement patterns, or activity frequency, these systems risk reducing complex human contributions to mechanical indicators of productivity. Similarly, the discussion expands how algorithmic systems alter not only managerial control but also employee agency, team dynamics, and the conditions for innovation. The case of Uber, among others, highlights the emergence of new forms of digital resistance and collective adaptation, phenomena that challenge deterministic narratives about technology and control.

By situating algorithmic governance within the broader perspectives of STS and digital Taylorism, it is argued that these systems are not merely technical tools but also carriers of organizational values and power structures. Therefore, their design and deployment must be guided by ethical commitments to inclusiveness, accountability, and human dignity. The ethical and social implications of data-driven managerial systems emphasize the need for organizations to align technological efficiency with fairness and human values \cite{Stark2024, KeeganMeijerink2025}.

Looking ahead, the growing integration of automated management tools in HRM calls for sustained interdisciplinary research into how employees perceive and respond to this shift in managerial authority. The collaboration between HCI and HRM offers a powerful foundation for reimagining algorithmic systems that not only optimize work but also support autonomy, creativity, and employee well-being. This requires moving beyond compliance and risk management toward a proactive, ethical, and human-centered approach to technological innovation.

However, much remains to be understood. Future research could explore the long-term impacts of automated managerial systems on employee well-being through longitudinal studies, assessing how these systems shape employee satisfaction and mental health over time. Comparative studies across industries could reveal how algorithmic systems are adopted and adapted in different organizational contexts, helping uncover sector-specific challenges and solutions. Furthermore, experimental designs could test interventions aimed at mitigating the negative impacts of algorithmic systems, such as incorporating more human oversight or promoting transparency in decision-making. These avenues are critical to ensuring such systems drive efficiency while aligning with broader social values and human rights.

Ultimately, the future of algorithmic governance hinges on how employees perceive and respond to the shift from human managers to digital control, and how organizations balance the operational benefits of automation with the fundamental need for fairness, transparency, and human agency in the workplace. To achieve this, organizations must embrace a proactive approach that integrates ethical considerations and fosters ongoing dialogue among technology developers, managers, and employees. In doing so, they can harness the transformative potential of algorithmic systems while supporting employee autonomy and agency. Workplaces must be designed not only for efficiency, but also for equity, justice, and resilience, where technological innovation and human talent co-evolve.

As automated management moves from an experimental edge case to a mainstream practice \cite{KeeganMeijerink2025}, organizations must confront both operational and ethical challenges. This calls for a reevaluation of how technological systems can align efficiency goals with norms of organizational justice, mitigating evaluative opacity while supporting employee agency \cite{Stark2024}.

\section*{Declaration of Interest}
The author declares that no known conflicts of interest are associated with this publication. This research received no specific grant from funding agencies in the public, commercial, or not-for-profit sectors.

\section*{Acknowledgments}
The author thanks the IT University of Copenhagen, Olayide Oladokun, and Christian Code for their valuable feedback and discussions that helped improve this work.

\end{document}